\documentclass[rnote]{aa} 
\usepackage{graphicx}
\usepackage{natbib}
\bibpunct{(}{)}{;}{a}{}{,}
\usepackage{txfonts}
\def\FeH{[{\rm Fe}$/${\rm H}]}
\def\FeHsun{$[{\rm Fe}$/${\rm H}]_{\odot}$}

\begin{document}
   \title{On the spectral resolution of the MILES stellar library}


\author{A. Beifiori, C. Maraston, D. Thomas, J.  Johansson} 
         
   \institute{Institute of Cosmology and Gravitation, University of
     Portsmouth, Dennis Sciama Bldg, Burnaby Road, Portsmouth, PO1
     3FX, UK \\
SEPNET, South East Physics Network, UK\\
              \email{alessandra.beifiori@port.ac.uk}
                      }

\authorrunning{Beifiori et al.}

   \date{Received December 14, 2010; accepted ...}


\abstract  
 {Empirical stellar libraries are extensively used to extract
  stellar kinematics in galaxies and to build stellar population
  models. An accurate knowledge of the spectral resolution of these
  libraries is critical to avoid propagation errors and uncertain
  estimates of the intrinsic stellar velocity dispersion of galaxies.}
{In this research note we re-assess the spectral resolution of the
  MILES stellar library and of the stellar population models based on
  it. This exercise was performed, because of a recent controversy
  over the exact MILES resolution.}
{We perform our test through the comparison of MILES stellar
spectra with three different sets of higher-resolution templates, one
fully theoretical - the MARCS library - and two empirical ones, namely
the Indo-U.S. and ELODIE v3.1 libraries. The theoretical template has
a well-defined very high (R=20000) resolution. Hence errors on this
theoretical value do not affect our conclusions. Our approach based on
the MARCS library was crucial to constrain the values of the
resolution also for the other two empirical templates.}
{We find that the MILES resolution has previously been slightly
  overestimated. We derive a new spectral resolution of 2.54 \AA\
  FWHM, instead of the nominal 2.3 \AA. The reason for this difference
  is due to an overestimation of the resolution for the
  Indo-U.S. library that was previously used for estimates of the
  MILES resolution. For the Indo-U.S. we obtain a new value of 1.35
  \AA\ FWHM. Most importantly, the results derived from the MARCS and
  ELODIE libraries are in very good agreement.}
{These results are important for users of the MILES spectra library
  and for further development of stellar population models aimed to
  obtain accurate stellar kinematics in galaxies.}


   \maketitle

%

\section{Introduction}
\label{sec:Intro}

Stellar libraries are a crucial ingredient of stellar population
models that predict the spectral energy distributions of stellar
populations
\citep[e.g.][]{Bruzual2003,Vazdekis1996,Maraston2009a,Vazdekis2010,Maraston2010}. The
latter are a key tool to analyse unresolved stellar populations such as
galaxies and extra-galactic star clusters. The spectral resolution of
a stellar population model is determined by the spectral resolution of
the input stellar library. The latter can be either theoretical or
empirical. As theoretical model atmosphere calculations are known to
suffer from incomplete line lists and continuum uncertainties
\citep{Korn2005,Thomas2010,Maraston2010}, empirical stellar libraries
became a complementary and widely used option to calculate stellar
population models
\citep{Vazdekis1999,Bruzual2003,Maraston2009b,Maraston2010}. The
spectral resolution of empirical libraries is fixed by the
instrumental resolution of the observations. An accurate assessment of
the underlying spectral resolution of both the input library and the
final stellar population model is essential for the proper use of such
models.

Several new empirical stellar libraries have been published in recent
years, e.g., ELODIE \citep{Prugniel2001}, STELIB \citep{LeBorgne2003}
and MILES \citep{SanchezBlazquez2006}. The new stellar population
models of \citet{Maraston2010} include all these three libraries,
while the new model of Lick absorption-line indices by
\citet{Thomas2010}, \citet{Johansson2010} is based on MILES. In both
\citet{Maraston2010} and \citet{Thomas2010} it has been pointed out
that the stellar population spectral resolution of MILES appears to be
somewhat coarser than generally assumed. It has been found that velocity
dispersions of galaxies from the Sloan Digital Sky Survey (SDSS) data
base \citep{York2000} derived using the MILES-based templates of
\citet{Maraston2010} agree well with the values from the SDSS-MPA/JHU
data base \citep{Kauffmann2003,Tremonti2004}. This indicates that
stellar population models based on MILES are close to
the SDSS spectral resolution ($R\sim 1800-2000$ at $5000\;$\AA), hence
somewhat coarser than the $2.3\;$\AA\ FWHM stated in
\citet{SanchezBlazquez2006} and \citet{Vazdekis2010}.

The aim of this research note is to assess the spectral resolution of
the MILES stellar library through direct comparison with various other
empirical and theoretical stellar libraries at higher spectral
resolution. Since a stellar population model may further dilute the
nominal spectral resolution of the input library owing to
uncertainties in radial velocities of the stars \citep{MacArthur2009},
we additionally assess the spectral resolution of the MILES-based
\citet{Maraston2010} model.


\section{Template stellar libraries}
\label{sec:MILES library}

The MILES library \citep{SanchezBlazquez2006} comprises 985 stars for
a wide range of evolutionary stages and metallicities. It covers a
wavelength range from $3500-7428\;$\AA\ at a {\it nominal} spectral
resolution of $2.3\;$\AA\ FWHM. This value has been determined in
\citet{SanchezBlazquez2006} through the comparison with the
higher-resolution empirical library Indo-U.S.\ \citep{Valdes2004}.

The result from this evaluation is certainly only as robust as the
spectral resolution of the adopted high-resolution library. Therefore,
in this research note we select three different stellar libraries,
both theoretical and empirical, with spectral resolutions higher than
the one of MILES, with the aim at checking the robustness of this value.

The adopted libraries are Indo-U.S., MARCS and ELODIE v3.1. The key to
our analysis is that MARCS is a theoretical library therefore with a
well-defined spectral resolution. We summarise below the main
characteristics of these libraries.

\begin{itemize}

\item {\it Indo-U.S. empirical library}: consists of 1273
  stars spectra covering the wavelength range between 3460-9464
  \AA\ with a {\it nominal} resolution of $1\;$\AA\ FWHM
  \citep{Valdes2004}. The library has a broad coverage of the
  stellar atmospheric parameters effective temperature, surface
  gravity and metallicity.

\item {\it MARCS theoretical library}: contains
  high-resolution (R=20000) theoretical spectra and covers a large
  spectral range from 1300 \AA\ to 20 $\mu$m \citet{Gustafsson2008}.
  Effective temperatures range between 2500-8000 $K$
  (at solar metallicity), surface gravities between -1.0 and 5.5 and
  metallicities between -5.0 and +1.0.

\item {\it ELODIE v.3.1 empirical library}: this version
  \citep{Prugniel2007} is an updated release of the
  original ELODIE library \citep{Prugniel2001} and includes
  1962 spectra of 1388 stars. ELODIE v3.1 covers a wavelength range
  between $3950-6739\;$\AA\ and provides a large coverage of
  atmospheric parameters. It is given at two resolutions, $R=42000$
  and $R=10000$. We decided to select the
  R=42000 (LH\_ELODIE), which is more adequate for a robust test of
  the MILES library resolution.

\end{itemize}

\section{Method}
\label{sec:method_results}

To extract the resolution of the MILES spectra we follow a similar procedure
as in \citet{SanchezBlazquez2006}.
We first divide each spectrum in 11 regions equally spaced in log
space to be able to assess the dependence of the resolution with
wavelength.
We then derive the broadening of each MILES star with respect to a library of
templates at higher resolution (Indo-U.S., MARCS, ELODIE v3.1) by using the
Penalized Pixel-fitting method (pPXF) of \citet{Cappellari2004} and
taking into account the resolution of each template.
pPXF performs the fitting in the pixel scale between an observed
spectrum and a linear combination of templates. pPXF also allows the use of
additive and multiplicative Legendre polynomials to adjust the
continuum shape of the template to the one of the spectrum to be analysed.
We made some checks and found that the final results are
consistently good by adding or not these polynomials. Therefore for
the general test we decide not to use them.

The comparison between the results of two empirical libraries and a
theoretical library gives us indication not only on the resolution of
MILES but also on the actual resolution of the empirical
templates we use.
Hence, by using the same method, we additionally estimate the actual FWHM
of the Indo-U.S. library using as templates MARCS and ELODIE v3.1
libraries.
Note that the very high resolution of our templates - ELODIE v3.1
(R=42000) and MARCS (R=20000) - make us confident in our results since
any error in the nominal resolution has very little impact on the
final derived value.

\begin{figure}
   \centering 
   \includegraphics[width=\columnwidth]{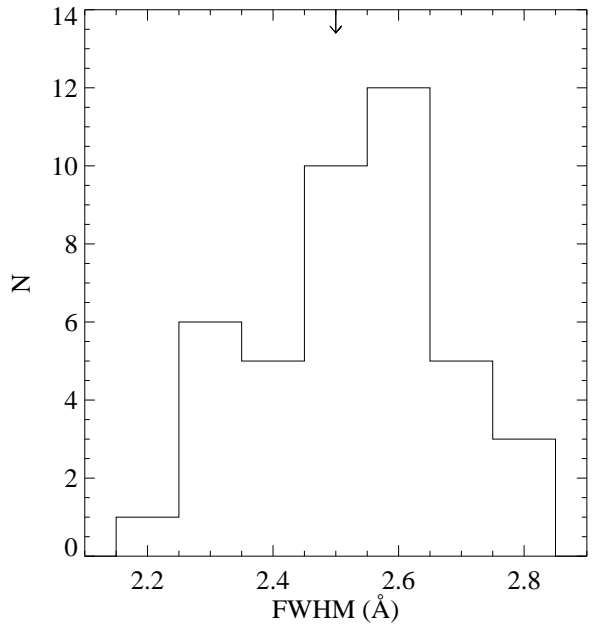}
   \caption{Distribution of FWHM for the 42 MILES stars fitted
     with the 456 ELODIE v3.1 stars for the wavelength bin from
     5160-5516 \AA.The median of this distribution is marked by an
     arrow.}
         \label{fig:Fig1}
   \end{figure}

\begin{figure}
   \centering 
   \includegraphics[width=\columnwidth]{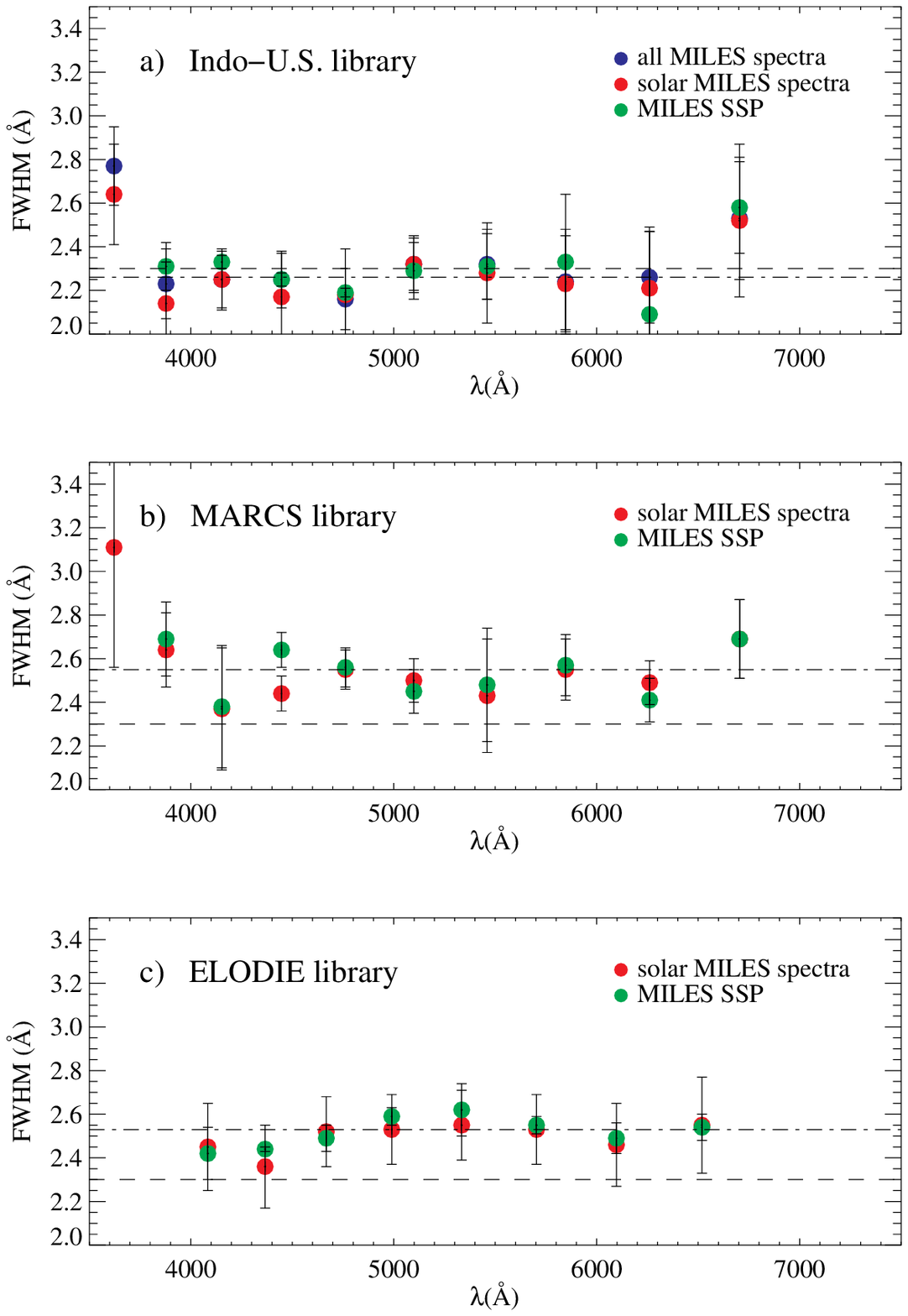}
   \caption{FWHM as a function of wavelength for the
     three different templates, {\em panel a)} Indo-U.S., {\em panel
       b)} MARCS, and {\em panel c)} ELODIE v.3.1 library. Different
     colours identify the MILES library (blue), the subsample of MILES
     stars with solar metallicity (red), the MILES SSP from
     \citet{Maraston2010} (green). The dashed line represents the
     nominal FWHM =2.3\AA\ \citep{SanchezBlazquez2006}, whereas the
     dotted-dashed line represents the median of the FWHM obtained here. 
     In the first wavelength bin of {\em panel a)} and
     {\em panel b)} the FWHM of SSP are not shown because out from the
     wavelength range (see Maraston \& Stromback for details).}
         \label{fig:Fig2}
   \end{figure}

After obtaining the values of the broadening for each MILES star we
derive the median of FWHMs to avoid being dominated by spurious values.
Errors for each wavelength bin
were estimated as the standard deviation of the FWHMs.  This
estimation is validated by the fact that the typical distribution of
FWHM is close to a Gaussian (see Fig.\ref{fig:Fig1} for one example).
In Fig.~\ref{fig:Fig2} we show the results for each
wavelength bin and for each template library as described below.

With the same procedure we derived the resolution of the MILES-based
SSP of \citet{Maraston2010} by using a representative subsample of
them covering the whole age range - 6.5 Myr, 1Gyr and 10 Gyr. This is
important as the fractional contribution of dwarfs and giants to the
integrated stellar population spectrum changes with age. We use solar
metallicity models with a Salpeter IMF.
In all tests we decided to exclude the last wavelength bin at
$>7000$\AA\ because the fit was not as reliable as in
the other bins for none of the MILES stars.

   \begin{figure}
   \centering 
   \includegraphics[width=\columnwidth]{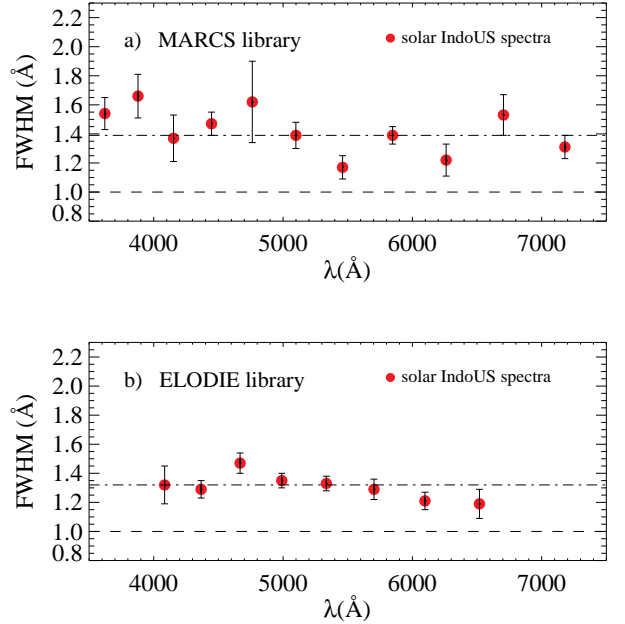}
   \caption{FWHM of Indo-U.S. library (red dots) as a function of wavelength
     for two different templates {\em panel a)} MARCS, and {\em
       panel b)} ELODIE v.3.1. In each plot the dashed line
     represents the nominal FWHM= 1\AA\ from \citet{Valdes2004},
     whereas the dotted-dashed line represents the median FWHM found here.}
         \label{fig:Fig3}
   \end{figure}

\section{Results}
\label{subsec:results}

In the following we describe the results for each template separately.
Note that in all cases we decided to remove the reddest wavelength
interval ($6937-7428\;$\AA) as we could not obtain a good fit to the
spectra, possibly due to the lack to strong absorption features to
constrain the fit.

\subsection{Template: Indo-US library at FWHM=1\AA}
\label{subsubsec:results_IndoUS}

As a first test we follow the same procedure of
\citet{SanchezBlazquez2006}.  
We first fitted each 985 MILES spectra with a linear combination of
Indo-U.S. spectra (1274 templates).  Then we applied the same
technique to a subsample of stars chosen among the MILES and
Indo-U.S. solar metallicity stars (\FeH=\FeHsun$\pm0.01$) finding 42
and 44 stars, respectively.  This could give indications about
possible bias in the selection of the stars. We did not find any bias
between the initial sample and the final one, which excludes template
mismatch and an insufficient number of templates.
In the following we will use only the solar metallicity subsample.
The results are shown in panel a) of Fig.~\ref{fig:Fig2} (see the
distribution of blue and red point in Fig.~\ref{fig:Fig2} panel a) for
differences between all MILES stars and the solar metallicity
subsample.).

Results are consistent for most wavelength regions. We found a lower
resolution in the first bin ranging from $3500-3748\;$\AA, possibly
due to the higher residuals in the fit we found in this wavelength
interval.
For the stellar population models in particular, this first bin gives
a relatively poor resolution of $4.31\pm 0.03\;$\AA\ FWHM, hence a
significantly lower spectral resolution. This wavelength bin is
therefore excluded from the analysis.

We also notice a difference between the resolution in the redder
and bluer part of the spectrum but with a variation that is negligible
within the errors.
The median FWHM derived by using Indo-U.S. is 2.26 $\pm$ 0.08 \AA\ and
2.31 $\pm$ 0.11 \AA\ for stars and stellar population models,
respectively.  The error was derived as the standard deviation of the
10 values.  The median instead of the mean was used in order not to be
affected by outliers.

\subsection{Template: MARCS library at R=20000}
\label{subsubsec:results_MARCS}

We use all solar metallicity MILES stars and combine them with the
theoretical MARCS library selecting the spectra at intervals of T=500
$K$~(92 spectra in total).
We follow the same procedure as above
and we found the same issue for the latest and first wavelength bin in
both the stellar spectra and integrated models.
The results are shown in panel b) of Fig.~\ref{fig:Fig2}.  The median
FWHM is 2.55 $\pm$ 0.14 \AA\ and 2.57 $\pm$ 0.13 \AA\ for stars and
stellar population models, respectively.

\subsection{Template: ELODIE v3.1 library at R=42000}
\label{subsubsec:results_ELODIE}

We fit MILES solar metallicity stars with ELODIE solar metallicity
templates (456 stars).  Since ELODIE covers a shorter wavelength range
(3950 - 6739 \AA) than MILES (3500-7428 \AA) we select the MILES
spectra between 3950 - 6739 \AA.  We divided the spectra in 8
wavelength regions instead of 11 to match the bins we used in the two
tests described above.
Note that even though the use of ELODIE limits our test to a shorter
wavelength range, we are still able to cover the most important
spectral features to extract the stellar kinematics.
The results are shown in panel c) of Fig.~\ref{fig:Fig2}.
The median FWHM we derived by using ELODIE v3.1 is 2.53 $\pm$
0.06 \AA\ and 2.54 $\pm$ 0.07 \AA\ for stars and stellar population models,
respectively.

\subsection{Comparison and discussion}
The results we obtained with both the ELODIE and MARCS templates agree
well with each other and indicate a resolution of $\sim 2.54 \pm0.08$
\AA\ FWHM for both MILES stars and the stellar population models based
on those stars.

This value is smaller than the nominal value for the MILES
resolution of $2.3\;$\AA\ FWHM. The resolution obtained by means of
the Indo-U.S. library, instead, is consistent with the latter.
This implies that the different resolution for MILES found by
\citet{SanchezBlazquez2006} comes from the different resolution of the
underlying template - Indo-U.S. - that was used to derive the
resolution. The triple test presented here suggests that the library
Indo-U.S. has a coarser spectral resolution than previously thought.

To test this conclusion we re-evaluate the spectral resolution of the
Indo-U.S.\ library following the same procedure as described above. We
fitted the solar metallicity Indo-U.S. stars with the two high
resolution libraries MARCS and ELODIE.  The results are shown in
Fig.~\ref{fig:Fig3}. The resulting median resolution of the
Indo-U.S.\ library turns out to be $1.39\pm 0.17\;$\AA\ FWHM and
$1.32\pm 0.09\;$\AA\ for the MARCS and ELODIE templates,
respectively. This is significantly lower than the $1\;$\AA\ FWHM
value specified in \citet{Valdes2004}. We conclude that the spectral
resolution derived by \citet{SanchezBlazquez2006} for MILES was too
high, mainly because the spectral resolution of the library used for
the fit was overestimated. The true spectral resolutions of the MILES
and Indo-U.S. libraries are $2.54 \pm 0.08\;$\AA\ FWHM and $1.35\pm
0.07\;$\AA\ FWHM, respectively.

\section{Conclusions}
\label{sec:conclusion}
Over the last few years, empirical stellar libraries have been
extensively used to build stellar population models in combination
with theoretical spectra.  A very accurate knowledge of the spectral
resolution is fundamental to avoid propagation errors and wrong
estimation of the intrinsic stellar velocity dispersion of galaxies.

In this research note we report results from tests aimed at verifying
the spectral resolution of MILES library and the SSP models based on
them.
This exercise was done because in \citet{Maraston2010} and
\citet{Thomas2010} it was found that the spectral resolution of stellar
population models based on MILES appeared to be somewhat coarser than
generally assumed.

We base our test on the comparison of MILES stars with three different
sets of templates, one theoretical, from the high-resolution MARCS
library, and two empirical, the Indo-U.S. library initially used for
assessing the spectral resolution of MILES by
\citep{SanchezBlazquez2006} and ELODIE v3.1.  The key to our analysis is
that the theoretical template has a very high and well-defined
resolution, implying that any error on this theoretical value would
not affect our conclusion. The MARCS library was crucial to constrain
the values of the resolution also for the other two empirical
templates.

The results we obtained with both ELODIE and MARCS templates agree
very well with each other and indicate a resolution of $\sim 2.54
\pm 0.08$ \AA\ FWHM for both the MILES stars and the stellar population
models based on those stars. This is somewhat smaller than the nominal
value for the MILES resolution of $2.3\;$\AA\ FWHM.
We found that the difference is due to the uncertainties on the
Indo-U.S. library FWHM. By applying the same method and using ELODIE
and MARCS as templates for Indo-US, we find a resulting median
resolution of $1.35\pm 0.07\;$\AA\ FWHM. This implies a
significantly lower spectral resolution than the $1\;$\AA\ FWHM
specified in \citet{Valdes2004}, which has propagated into the
derivation of the MILES spectral resolution.

\begin{acknowledgements}
We are grateful to Harald Kuntschner, Patricia
S{\'a}nchez-Bl{\'a}zquez, and Jes{\'u}s Falc{\'o}n-Barroso for useful
discussions. AB, CM and JJ acknowledge support by the Marie Curie
Excellence Team Grant UniMass (PI C. Maraston) MEXT-CT-2006-042754 of
the Training and Mobility of Researchers programme financed by the
European Community.
\end{acknowledgements}

\end{document}